\documentclass[12pt,A4paper]{sn-jnl}% Default
\usepackage[version=4]{mhchem}
\usepackage{subcaption}
\usepackage{graphicx}
\makeatletter \providecommand\barcirc{\mathpalette\@barred\circ}
\def\@barred#1#2{\ooalign{\hfil$#1-$\hfil\cr\hfil$#1#2$\hfil\cr}}
\newcommand\stst{^{\protect\barcirc}}
\makeatother
\jyear{2022}%

\begin{document} %%%%%%%%%%%%%%%%%%%%%%%%%%%%%%%%%%%%%%%%%%%%%%%%%%%%

\title[Peptide Bonds in the ISM]{Peptide Bonds in the Interstellar Medium:\\Facile Autocatalytic Formation from Nitriles on Water-Ice Grains}

\author*[1,2]{\fnm{Bouthe\"ina}\sur{ Kerkeni}}\email{boutheina.kerkeni@isamm.uma.tn}
\equalcont{These authors contributed equally to this work.}
\author[3]{\fnm{John M.} \sur{Simmie}}\email{john.simmie@nuigalway.ie}
\equalcont{These authors contributed equally to this work.}

\affil[1]{\orgdiv{ISAMM},  \city{Universit\'e de la Manouba}, \country{Tunisia 2010}}
\affil[2]{\orgdiv{Facult\'e des Sciences de Tunis}, \orgname{Laboratoire de Physique de la Mati\`ere Condens\'ee}, \city{Universit\'e Tunis el Manar}, \country{Tunisia 2092}}

\affil[3]{\orgdiv{School of Chemistry}, \orgname{University of Galway},  \country{Ireland  H91 TK33}}

\abstract{A recent suggestion that acetamide, \ce{CH3C(O)NH2}, could be readily formed on water-ice grains by the acid induced addition of water across the \ce{CN} bond is now shown to be valid. Computational modelling of the reaction between \ce{R-CN} (R = H, \ce{CH3}) and a cluster of 32 molecules of water and one \ce{H3O+} proceeds auto-catalytically to form firstly a hydroxy imine \ce{R-C(OH)=NH} and secondly an amide \ce{R-C(O)NH2}. Quantum mechanical tunnelling, computed from small-curvature estimates, plays a key role in the rates of these reactions.  This work represents the first credible effort to show how amides can be formed from abundant substrates, namely nitriles and water, reacting on a water-ice cluster containing catalytic amounts of hydrons in the interstellar medium with consequential implications towards the origins of life.}

\keywords{amides, ISM, nitriles, quantum chemistry}



\maketitle

\section{Introduction}\label{sec1}
The peptide bond, \ce{-C(O)NH-}, found in amides connects amino acids to peptides --- of paramount importance to present day life on Earth.  Unsurprisingly, how, when and where peptide bond formation arose is of immediate interest in prebiotic astrochemistry, tackling that challenging question: the origins of life \cite{kolesnikova, ligterink}.

The simplest amides, formamide \ce{HC(O)NH2}, and acetamide, \ce{CH3C(O)NH2}, are common constituents of star-forming regions in our galaxy \cite{adande,mcguire} but apparently propionamide, \ce{CH3CH2C(O)NH2}, is not \cite{caden}. Adande et al. \cite{adande} have shown, based on their observations of formamide towards star-forming regions of dense molecular clouds, that the compound could have been brought to Earth by exogenous delivery in substantial amounts of $\sim 0.18$ mmol m$^{-2}$ in a single impact.

The formation routes to formamide are still unclear; some have suggested gas-phase pathways via formaldehyde and amidogen \cite{barone,skouteris}:
$$\ce{H2CO + NH2^. -> HCONH2 + H^.}$$
but disputed by Song and K\"{a}stner \cite{song} and convincingly refuted by Douglas et al. \cite{douglas} or surface reactions \cite{fedoseev} by the hydrogenation of isocyanic acid, \ce{HNCO}, and on carbon monoxide--ammonia ices \cite{jones}:
\begin{eqnarray}
\ce{NH3 -> NH2^. + H^.} \\ 
\ce{H^. + CO ->  HCO^.}\\ 
\ce{HCO^. + NH2^. -> HC(O)NH2}
\end{eqnarray}
and more recently by metal-ion mediated substitution reactions \cite{thripati}: 
$$ \ce{HC(O)X + NH3 ->[M+] HC(O)NH2 + HX}$$
where \ce{M = Na+, K+, Mg+, Mg++, Al+} and X = \ce{H, OH, CH2OH} but the evidence for all these is underwhelming.  A comprehensive summary has been given recently by Chuang et al. \cite{chuang} during the course of their laboratory work on the formation of formamide in water- and carbon monoxide-rich water ices with ammonia as the substrate. VUV irradiation of water-rich and CO-rich ammonia ices at 10~K with compositions \ce{H2O{:}CO{:}NH3}=10:5:1, \ce{CO{:}NH3}=4:1 and \ce{CO{:}NH3}=0.6:1 have shown that formamide is preferentially formed, although mechanistically the situation is complicated with no clear indication as to actual formation routes \cite{chuang}.
Indirect evidence for the formation routes of formamide based on the stratified distribution of the molecules \ce{HNCO} and \ce{H2CO}, putative parents of \ce{HCONH2}, in the atmosphere of the HH 212 protostellar disk appear to rule out \ce{HNCO} as a parent \cite{codella}.

Studies of comets have indicated that the early Solar Nebula had nitriles (cyanides) such as \ce{HCN} and \ce{CH3CN} in abundance \cite{cordiner,loomis}. 
It is currently assumed that reactions in the bulk or on the surface of water-ice grains are the most likely formation routes for complex organic molecules which are then liberated into the gas-phase by UV irradiation, electron and cosmic ray bombardment, via thermal shocks or grain--grain collisions \cite{colzi,kalv,mini}.

In a computational study of the direct reaction \ce{HC#N + H2O -> HC(OH)=NH} neither the presence of a second \ce{H2O} as a catalyst, or as a spectator or as a reactant was sufficient to reduce the high barriers encountered which thereby rule out the possibility of it occurring in cold molecular clouds \cite{darla}.
The most pertinent previous work simulated a 33 \ce{H2O} molecule cluster and showed that \ce{HC#N} cannot react to \ce{HC(O)NH2} under interstellar ice conditions because of large energy barriers.  Any reactivity that was feasible proceeded through the \ce{CN^.} radical \cite{rimola}.

UV photolysis of water methyl cyanide ices at 20~K (\ce{H2O$:$CH3CN} = 20:1) does give rise to the formation of \ce{CH3C(O)NH2}, and its isomer  N-methyl formamide \ce{CH3NHCHO}, but many other products as well \cite{bulak}.

A consideration of the various suggested reactions in the literature  led one of us to suggest that \ce{H3O+} induced water addition to nitriles on water-ice grains was likely to provide the most probable route \cite{simmie}.  It is known that \ce{H3O+} exists in the ISM and even in other galaxies, that it reacts in water-ices  and that it has the potential to drive subsequent reaction \cite{wootten,tak,moon,lee}. Woon has very recently reviewed what is known about cation--ice reactions from quantum chemical cluster studies, highlighting the novel and more efficient pathways,  vis-\`a-vis the gas-phase, that \ce{HCO+}, \ce{CH3+} and \ce{C+} (but not \ce{H3O+}) undergo and appealed for experimental confirmation \cite{woon}.  Laboratory experiments have shown that bombardment of water ice samples on a copper substrate at 10~K yields a number of secondary ions including  $ \ce{(H2O)_n.H+}$ with $n=1\to 8$,  although  these ions are perhaps more accurately denoted as  $\ce{(H2O)_{n-1}.H3O+}$ \cite{martinez}.  The production efficiency is much lower when crystalline ice is bombarded in comparison to amorphous ice.

The first step in the proposed addition of water is the protonation at nitrogen of the nitrile to form \ce{RC#NH+}, and in the case of HCN to form imino methylium, \ce{HC#NH+}; this species has been widely detected in star-forming regions \cite{ziurys, schilke, quenard} and in Titan's atmosphere \cite{titan}. Although it had been characterised as a precursor of \ce{HC#N} in the most recent observations Fontani et al. show that it is formed from \ce{HC#N} or \ce{HC#N+} \cite{fontani}. Fundamental laboratory work to characterise the spectroscopic parameters for \ce{CH3C#NH+} have been carried out but interstellar searches have not so far been successful \cite{mari}.

A primary consideration for suggesting acidified amorphous water-ice is the high mobility of the proton through the lattice; the  transfer mechanism,  via Grotthus hops, takes place on a sub-picosecond time scale and with barriers of $\approx 1$ kJ mol$^{-1}$ --- these effectively increase the ``collision rate'' or encounter between reactant and \ce{H3O+} \cite{lee,hops}.              

The hydronium induced addition of water was seen as a two-step process with the first forming hydroxy imines or imidic acids \ce{RC(OH)=NH} and the second converting these to amides \ce{RC(O)NH2}; the latter reaction can occur either by intramolecular hydrogen-transfer via quantum-mechanical tunnelling or by the further protonation of the nitrogen-atom followed by deprotonation from the O-atom.

The actual formation of peptides as a by-product was observed \cite{krasno} after the deposition of C atoms onto a CO + \ce{NH3} ice at 10 K and subsequent warming to 300~K.  The authors argue that the initial reaction product is aminoketene, \ce{H2NCH=C=O}, which polymerises on warming yielding \ce{(-CH2-C(O)-NH-)$_n$} chains. While the experiments are compelling the conditions are somewhat artificial since they consider pure CO + \ce{NH3} ices with substantial quantities of carbon atoms.  The initial step \ce{H3N + C($^3\mathrm{P}_0$) -> H3N\bond{~}C} is highly exothermic, $\Delta _rH=-103$ kJ mol$^{-1}$, this is followed by a 1,2-H-transfer to \ce{H2N-\mbox{\"{C}}H} and finally reaction with CO to \ce{H2N-CH=C=O}.

Interestingly Canepa \cite{canepa} in considering the survival rates of glycine, \ce{NH2CH2C(O)OH}, embedded in micrometeorites undergoing atmospheric re-entry has shown that aminoketene, the product of dehydration, would also survive.

In this paper we focus on a solid-state chemistry formation mechanism of \ce{R-C(OH)=NH} and subsequently \ce{R-C(O)NH2}. We perform electronic structure investigations of the energetics  and also thermal rate constants calculations of this reaction mechanism. This information can then be used in astrochemical modelling and may prompt experimental laboratory confirmation.

\begin{figure}[h] 
\centering
\includegraphics[width=\textwidth,scale=0.2]{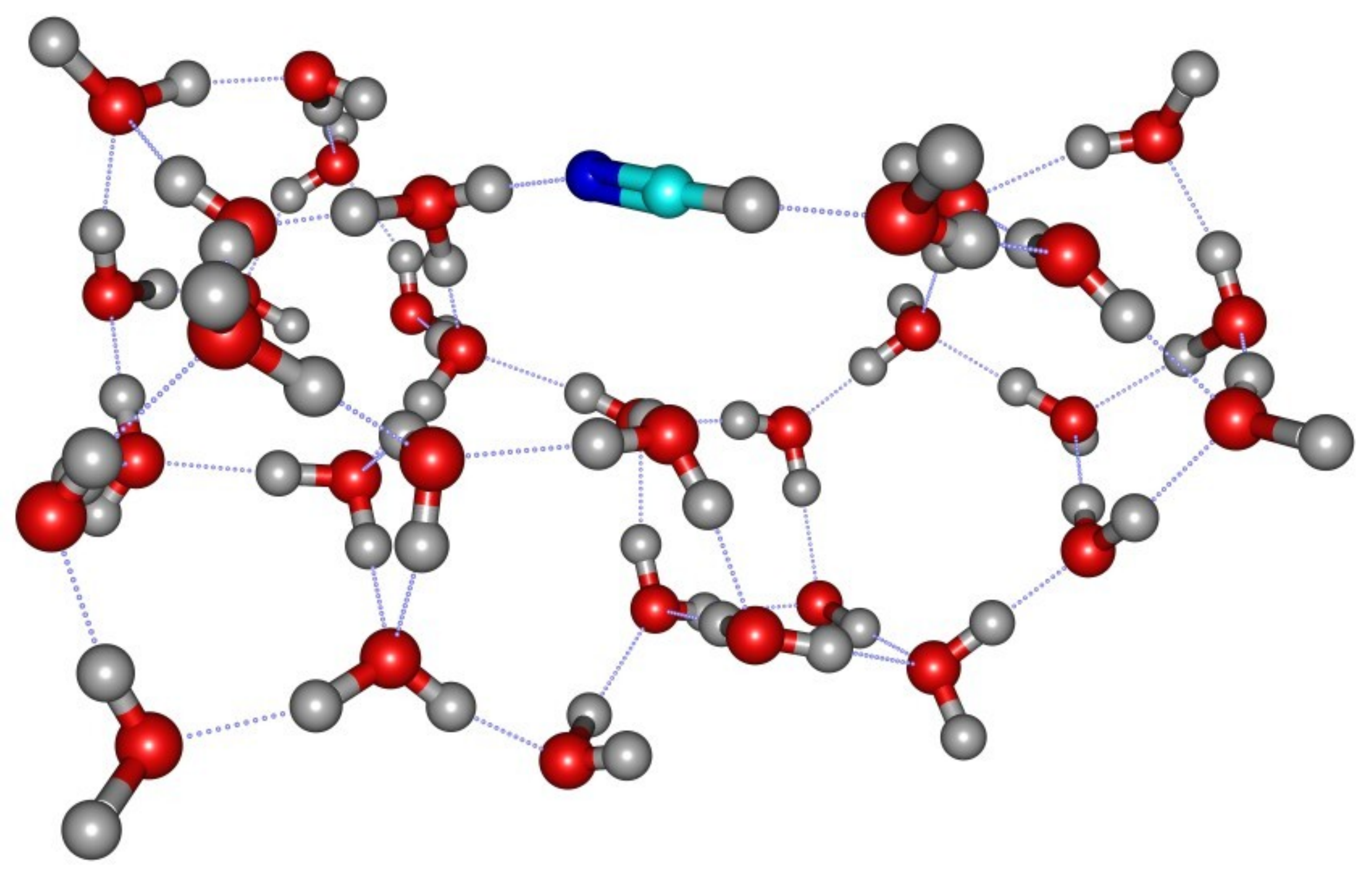}
\caption{HCN embedded in acidified water}
\label{water-hcn}
\end{figure}

\section{Methods}\label{sec11}
Calculations were performed with the application Gaussian \cite{gauss} and used the long-range dispersion corrected hybrid meta-GGA density functional $\omega$B97X-D together with the triple $\zeta$ basis set with added polarization and diffuse functions 6-311++G(d,p) \cite{wB97} with a factor of 0.96 was applied to scale the zero-point energy.

A system with thirty-two water molecules and one hydronium ion, \ce{H3O+}, with a total `volume' of $\approx 2,600$ \AA$^3$, was chosen together with one reactant, either \ce{HCN} or \ce{CH3CN}.  This choice represents a compromise between realistic ISM concentrations and computational effort Fig.~\ref{water-hcn}.

All structures were fully optimized and the harmonic frequencies computed using DFT.  
 Frequency calculations were performed in order to verify that all intermediates are true minima on the potential energy surface, and that all transition states exhibit a single imaginary frequency. We study all species in the reaction mechanism with the unrestricted $\omega$B97XD/6-311$++$G(d,p) model chemistry.
Gaussian 16 automatically includes an ultrafine integration grid in the DFT calculations in order to improve the accuracy of the results. The grid greatly enhances the accuracy at reasonable additional cost. 

The reaction paths are computed using the intrinsic reaction coordinate (IRC) methodology \cite{Hratchian2004,Hratchian2005} to confirm the identities of the reactants and products for every transition state. IRC calculations require initial force constants of the transition state. Then, the first and second order energy derivatives are obtained to calculate the projected harmonic vibrational frequencies along each reaction path. The minimum energy paths (MEPs) were computed using the Page--McIver integrator with a gradient step size of 0.1 $a_0$ \cite{page}.

Small curvature quantum mechanical and quantised-reactant-states tunnelling calculations \cite{sct,qrc} employed the PILGRIM application \cite{pil} for the computation of rate constants via transition state theory (TST)
and variational TST (VTST) necessitating calculations along the minimum energy path.

\section{Results}\label{sec2}
Prior to the first reaction steps we begin by considering a previously published \cite{Rim} cluster of 33 water molecules, \ce{[33(H2O)]},  to which a low energy hydron \ce{H+} is added in a highly exothermic process, $\Delta _rH(0 \mathrm{K})$ of $-1,065$ kJ mol$^{-1}$; this in turn can be dissipated throughout the cluster and/or serves to drive subsequent reactions.The only notable difference between the two clusters \ce{[33(H2O)]} and \ce{[33(H2O).(H+)]}, which is more realistically depicted as \ce{[32(H2O).(H3O+)]}, are the three additional vibrational modes, two \ce{O-H} asymmetric stretching vibrations near 2,400 cm$^{-1}$ and a characteristic symmetric at 2,800 cm$^{-1}$.  It is to this cluster that the reactants \ce{HCN} and \ce{CH3CN} are then added.

\subsection{First step: imidic acid formation}
As originally envisaged the hydrolysis of \ce{HC#N} proceeded in three distinct phases:
\begin{eqnarray}
\ce{HC#N + H3O+  -> HC=NH+ + H2O }\\
\ce{HC=NH+ + H2O -> HC(OH2)=NH+}\\
\ce{HC(OH2)=NH+ + H2O -> HC(OH)=NH + H3O+}
\end{eqnarray}
the exothermic first step $\Delta _rH\stst(0~K)=-20.5$ kJ mol$^{-1}$ simply a reflection of the larger proton affinity of \ce{HC#N} of 712.9 kJ mol$^{-1}$ vis-\`a-vis \ce{H2O} of 691.0 kJ mol$^{-1}$ \cite{hunter}.  In a water cluster of an additional 32 \ce{H2O} molecules however these later steps are elided since as \ce{H2O} adds to the C-atom it simultaneously transfers H to a neighbouring O-atom.
\begin{eqnarray}
\ce{HC#N + H3O+ + 32 H2O -> HC(OH)=NH + H3O+ + 31 H2O}
\end{eqnarray}

Overall reaction (4) is exothermic by $-80$ kJ mol$^{-1}$ which is considerably different to the gas-phase $\Delta _rH\stst(0 \mathrm{K})=-22.2$ kJ mol$^{-1}$ reflecting the tighter binding of the hydrolysed product in comparison to the reactant.

The barrier to reaction (1), that is protonation at the N-atom, in the case of HCN is 78.5 kJ mol$^{-1}$ and is even lower at 51.9 kJ mol$^{-1}$ for acetonitrile.

\begin{figure} 
\centering
\includegraphics[width=\textwidth,scale=0.2]{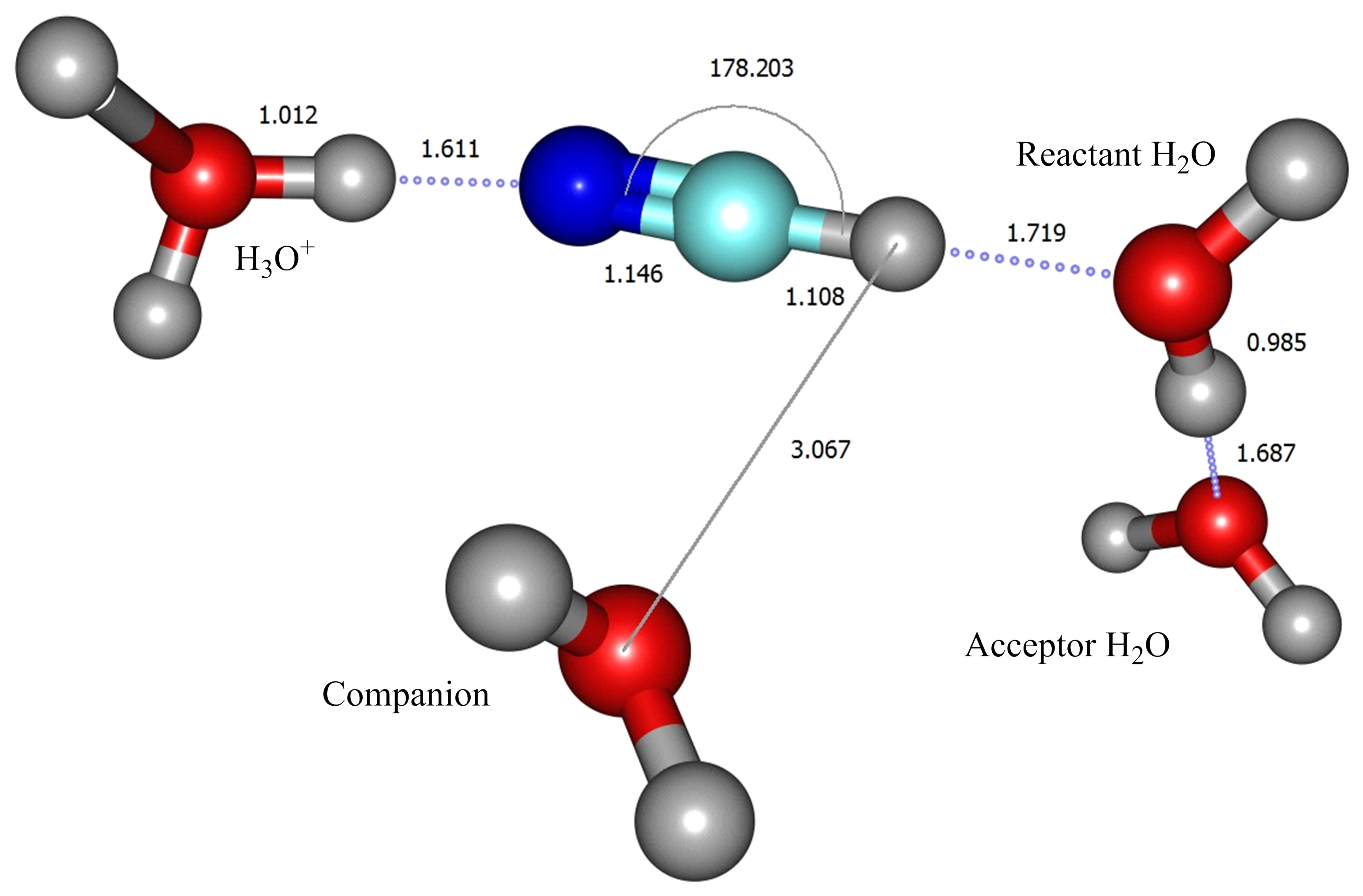}
\caption{Key elements of the reaction scheme for the first step}
\label{key}
\end{figure}

\begin{figure}
    \centering
    \begin{subfigure}[t]{0.45\textwidth}
    \includegraphics[width=\textwidth]{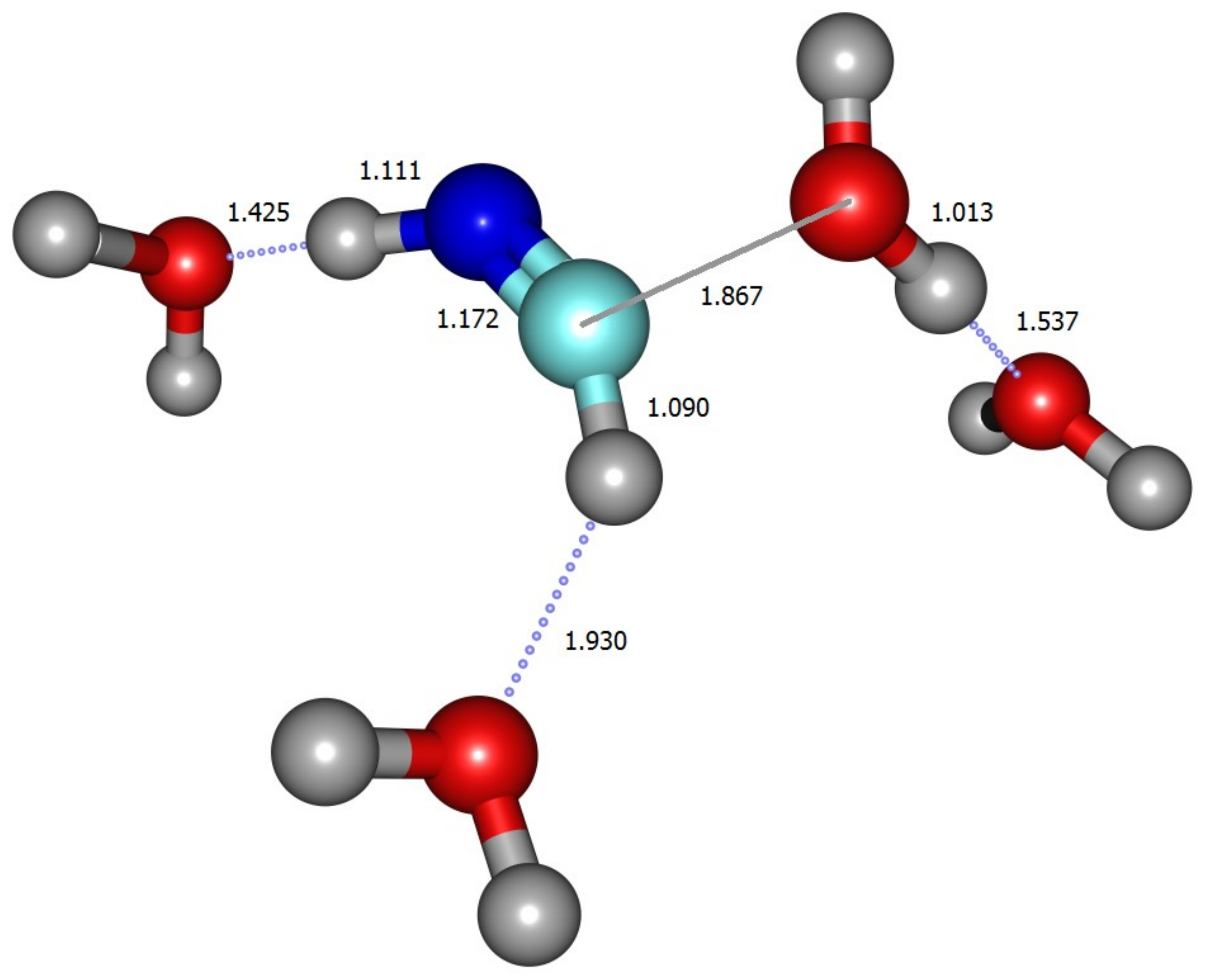}
    \caption{Transition state}
    \label{fig:first}
   \end{subfigure} 
Click to show the PDF

   \begin{subfigure}[t]{0.45\textwidth}
    \includegraphics[width=\textwidth]{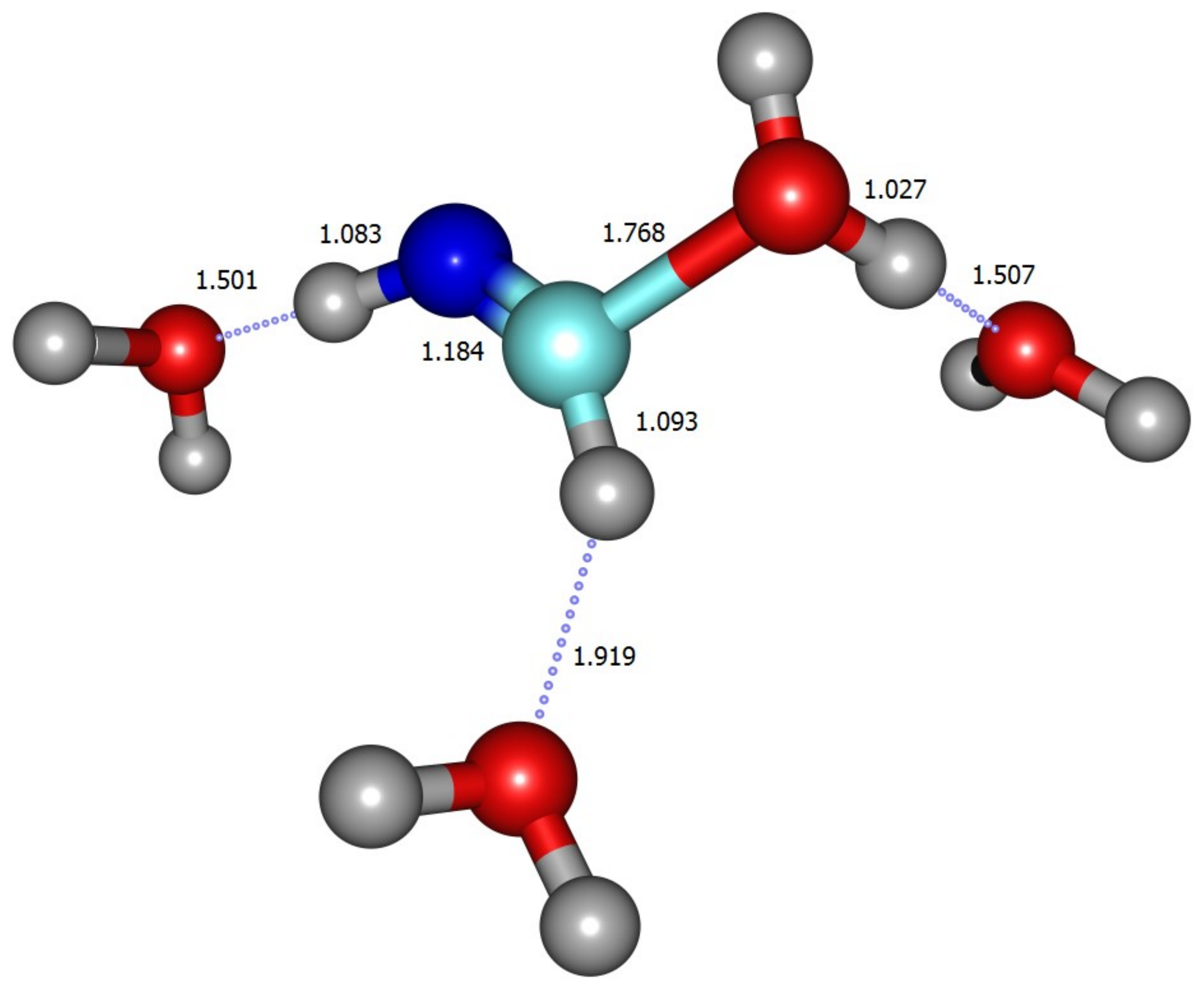}
    \caption{Addition of \ce{H2O}}
    \label{fig:second}
   \end{subfigure} 
\ \\
   \begin{subfigure}[t]{0.45\textwidth}
    \includegraphics[width=\textwidth]{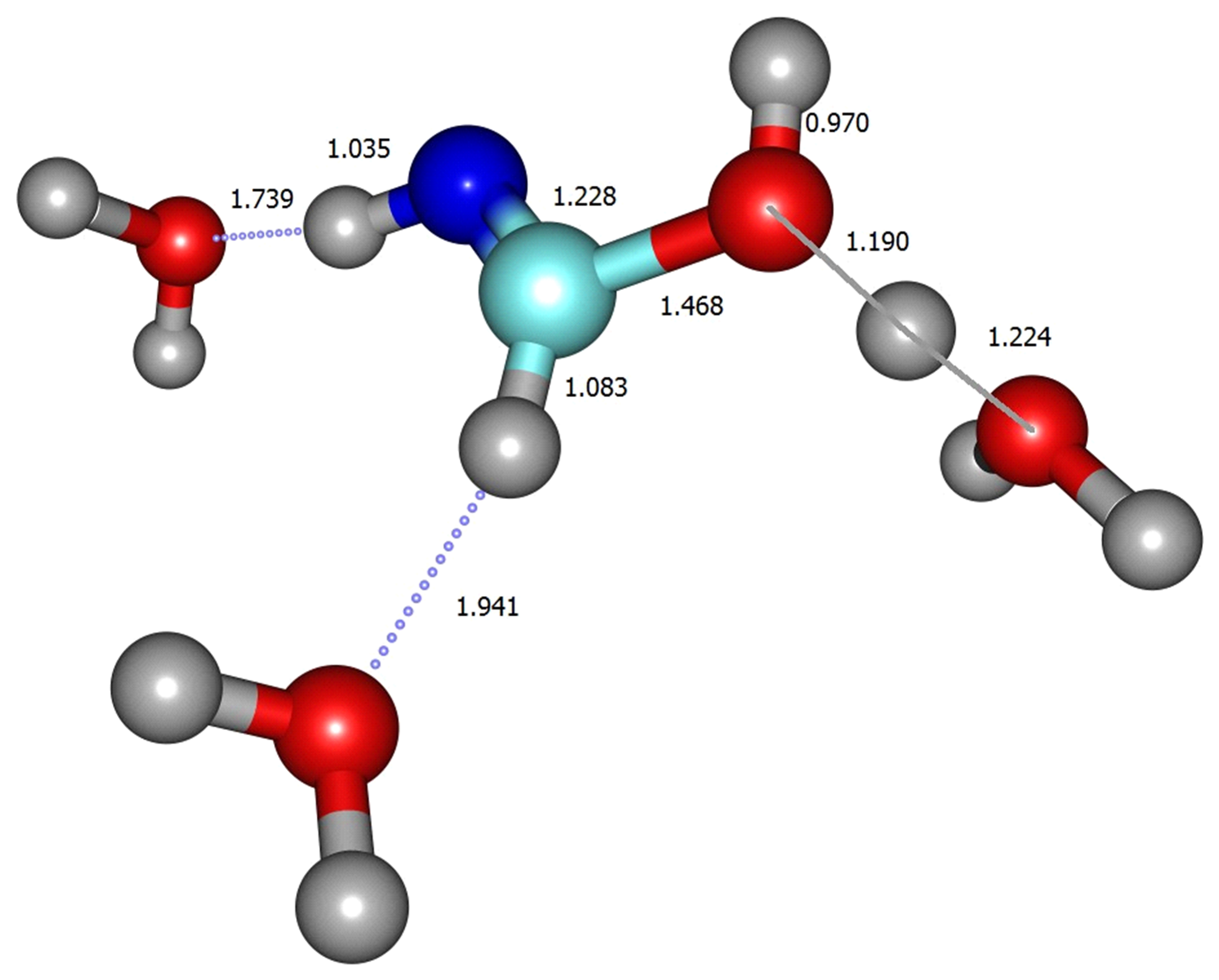} %% width=\textwidth,scale=0.45
    \caption{Transfer of \ce{H+}}
    \label{fig:third} 
   \end{subfigure} 
   \begin{subfigure}[t]{0.45\textwidth}
    \includegraphics[width=\textwidth]{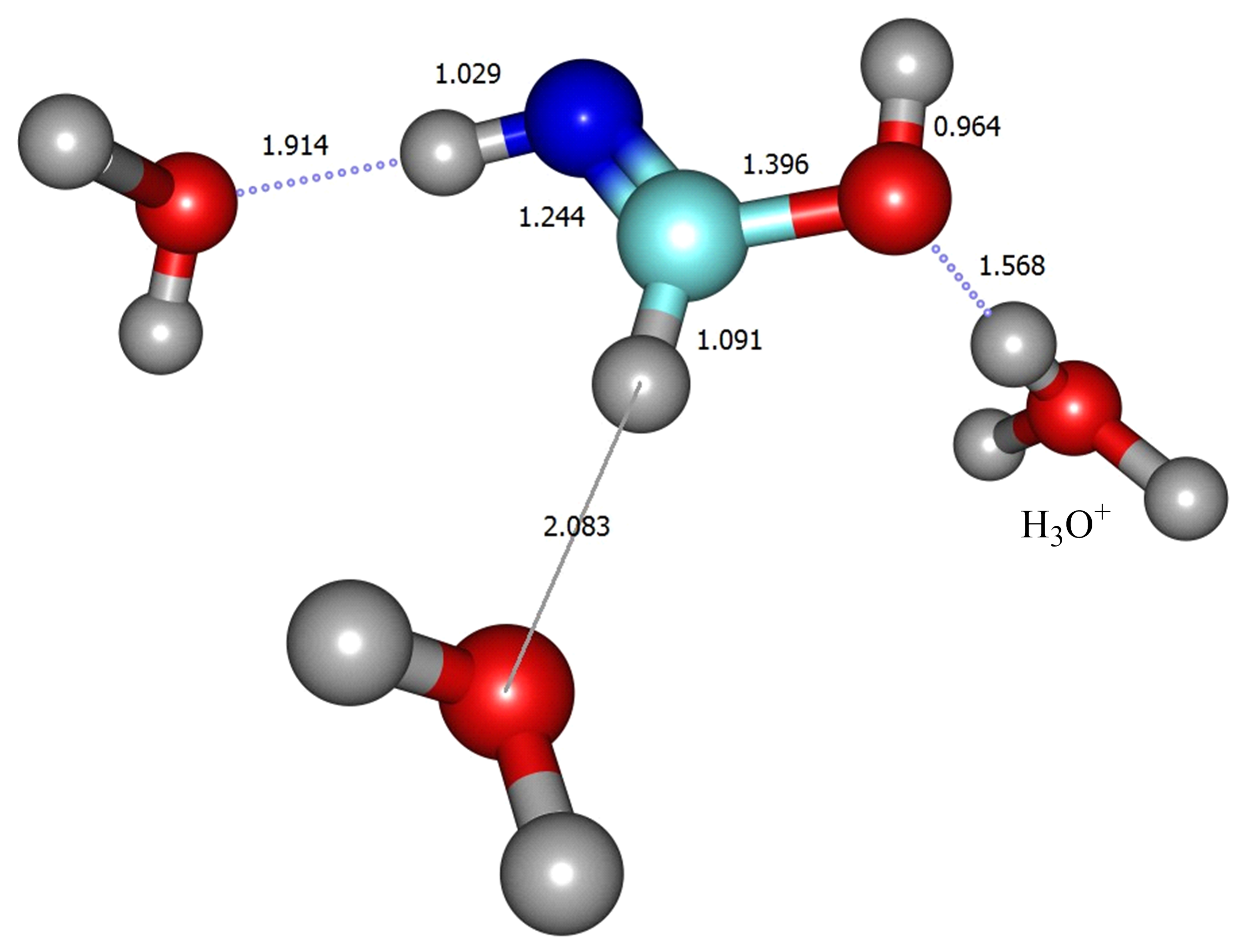}
    \caption{Final product}
    \label{fig:fourth} 
   \end{subfigure} 
 \caption{Structures along the IRC path for the first step}
 \label{snap}
\end{figure}

Although Fig.~\ref{water-hcn} shows the actual reaction structure, an oversimplified version, Fig.~\ref{key}, shows the parts played by four key water molecules; the first is the proton donor \ce{H3O+}, the second is the ``reactant'' which will attack the C-atom and also transfer a \ce{H+} to the ``acceptor'' water, whilst the ``companion'' \ce{H2O} is less involved but nevertheless stabilises the system through \ce{H\bond{...}OH2} hydrogen-bonding; snapshots along the IRC path are shown in Fig.~\ref{snap} to the final product, methanimidic acid in its (\textit{E,Z}) conformation with respect to the dihedrals $\angle$ OCNH and HOCN, respectively.

\begin{figure}
\centering
\includegraphics[width=\textwidth,scale=0.05]{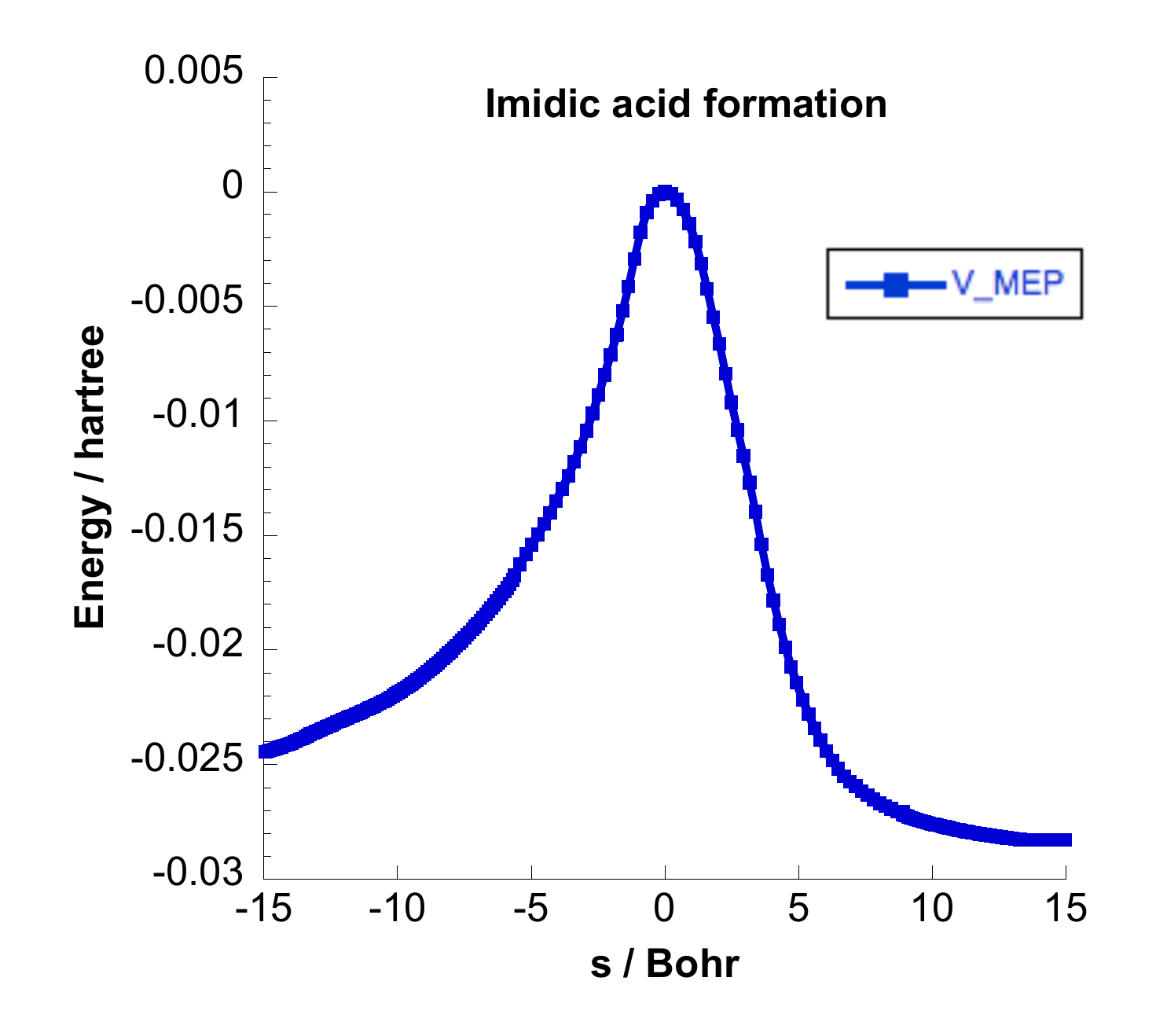}
\caption{Classical potential energies $V_{MEP}$ as functions of $s$ / Bohr for imidic acid formation in ice.}
\label{Figure:4}
\end{figure}

In Fig.~\ref{Figure:4} we plot the potential energy along the minimum energy path for the imidic acid formation given by reaction (4).

\subsection{Second step: amide formation}
We can distinguish  between two different routes from the hydroxy imines to the corresponding amides which can proceed intra-molecularly or inter-molecularly.

\subsubsection{Intramolecular route}
\begin{figure} 
\centering
\includegraphics[width=\textwidth,scale=0.2]{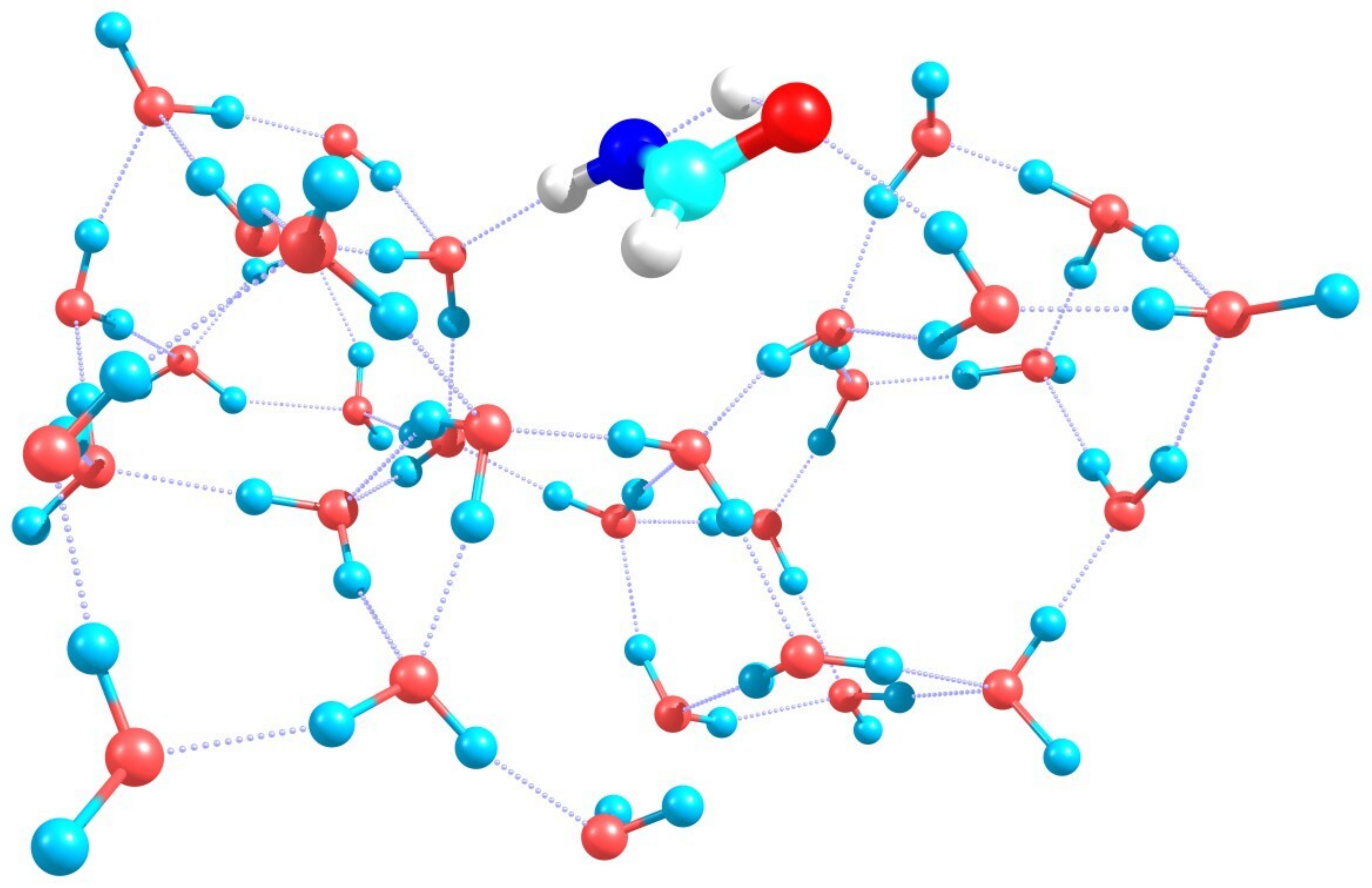}
\caption{Transition state for 1,3[H]-transfer; \ce{HC(OH)NH $\to$ HC(O)NH2}}
\label{13H}
\end{figure}

Once the hydroxy imines, methanimidic and ethanimidic acids, are formed then an intra-molecular 1,3-H-transfer leads to formamide, Fig.~\ref{13H}, or acetamide; however, the barriers are considerable
ranging from 136 or 128 kJ mol$^{-1}$ in the gas-phase to 169 and 142 kJ mol$^{-1}$ in this water-cluster; clearly, surmounting such barriers is unfeasible at temperatures much less than 300~K except by tunnelling. 

Note that the presence of additional water molecules makes very little difference to the energetics of the process in comparison to the gas-phase.  Exactly the same conclusion can be drawn from the gas-phase work of Darla et al. which showed in $\omega$B97xD/aug--cc-pVTZ calculations that even in the presence of a ``catalytic'' water molecule the 1,3[H]-transfer faces a barrier of 131 kJ mol$^{-1}$ or 132 kJ mol$^{-1}$ with the additional water present as a ``spectator'' \cite{darla}.

The additional water molecules and \ce{H3O+} in our system only marginally affect the 1,3[H]-transfer reaction; this is seen in the energetics, as discussed above, and also from the values of the imaginary frequencies which are gas-phase: 2,002 and 1,988 cm$^{-1}$ and cluster: 1,953 and 1,916 cm$^{-1}$ for \ce{HC(OH)NH $\leftrightarrow$ HC(O)NH2} and \ce{CH3C(OH)NH $\leftrightarrow$ CH3C(O)NH2} respectively.

\begin{figure} 
\centering
\includegraphics[width=\textwidth,scale=0.1]{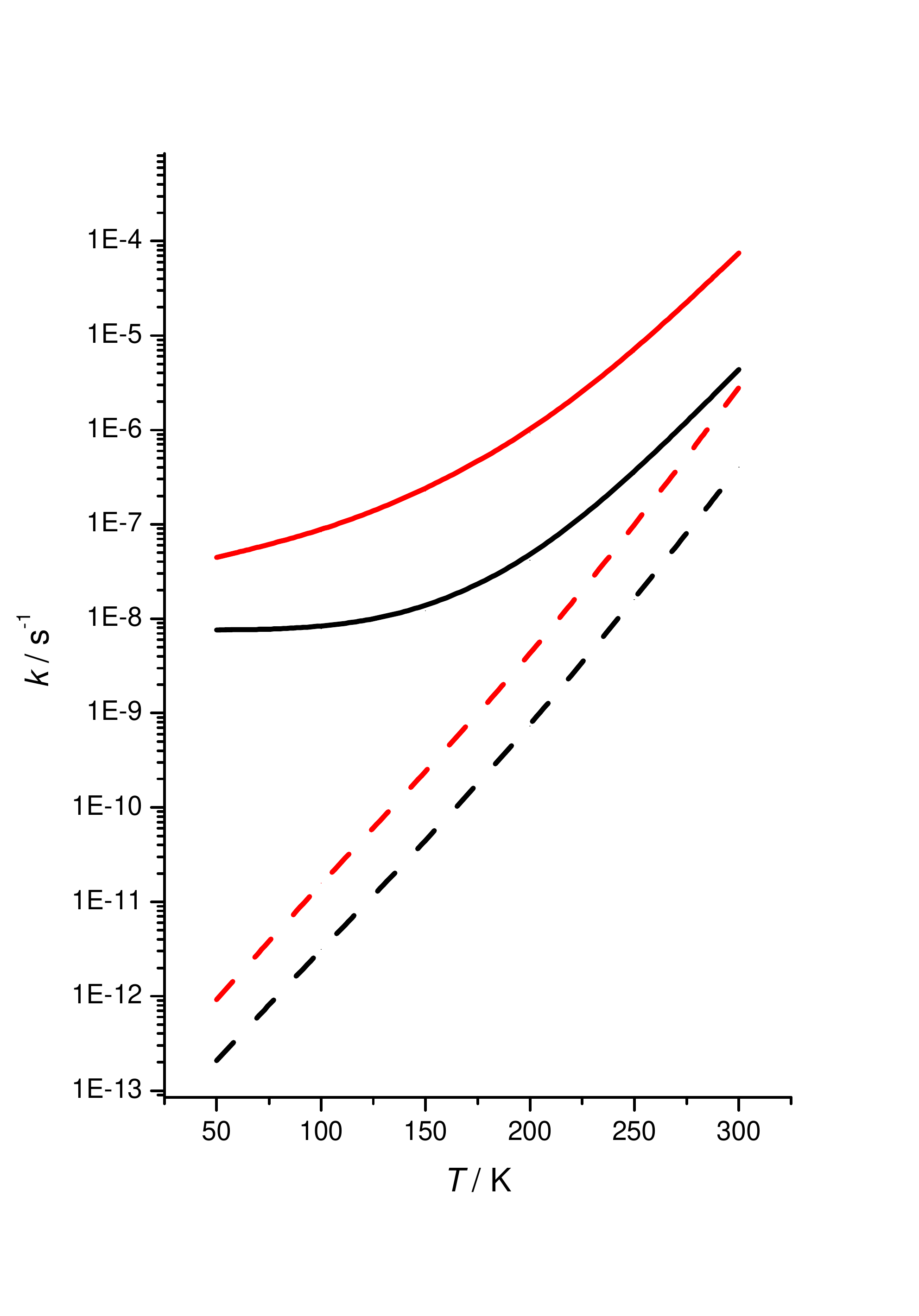}
\caption{Rate constants gas-phase isomerisation: Eckart versus small-curvature tunnelling. 
\textcolor{red}{\ce{CH3C(OH)NH}}, \ce{HC(OH)NH}.}
\label{comp}
\end{figure}

In the case of the gas-phase \ce{RC(OH)NH $\leftrightarrow$ RC(O)NH2} isomerisation reaction, PILGRIM \cite{pilgrim} calculations at B3LYP/cc-pVTZ incorporating small-curvature tunnelling yields rate constants, $k$, and half-lives, $\tau$, as shown in Table~\ref{tab:rates}; the ice-cluster values are probably not dis-similar.  At the lowest temperatures, \emph{here} $\leq 150$~K, quantised reactant states tunnelling is included. The barrier to reaction is higher at 131.5 kJ mol$^{-1}$ for \ce{HC(OH)NH2} than the 123.9 kJ mol$^{-1}$ for \ce{CH3C(OH)NH2} consequently the rate of isomerisation to the appropriate amide is faster for ethanimidic acid. These are substantially faster rates of isomerisation than our previous Multiwell values, Fig.~\ref{comp}, which were based on Eckart tunnelling \cite{simmie}; this renders the intermolecular route, discussed below, essentially redundant.

A not dis-similar situation is considered by Concepci\'on et al \cite{con} in their work on the origin of the (\emph{E}/\emph{Z}) isomer ratio of imines in the ISM; thus they show that the less-stable (\emph{E}) conformer of cyanomethanimine, \ce{RCH=NH} where R is a \ce{C#N} group, re-arranges to the (\emph{Z}) with dramatically increased rates over canonical transition-state theory values at temperatures of 250 K and below.  The variational effect is small and the faster rates are ascribed to quantum tunnelling; exactly the same as found \emph{here}.

\begin{table}[tbh]
    \centering
    \begin{tabular}{ccccc} 
                & \multicolumn{2}{c}{\ce{HC(OH)NH $\to$ HC(O)NH2} } & \multicolumn{2}{c}{\ce{CH3C(OH)NH $\to$ CH3C(O)NH2} }\\
     $T$ / K    &  $k$ / s$^{-1}$ & $\tau$ / days &  $k$ / s$^{-1}$ & $\tau$ / days\\ \hline
    \rule{0pt}{10pt} 50         &  $7.6 \times 10^{-09}$  & 1,060  &   $4.4 \times 10^{-08}$   &  180 \\
     100        &  $7.8 \times 10^{-09}$  & 1,028  &   $8.3 \times 10^{-08}$   &  100 \\  
     150        &  $1.2 \times 10^{-08}$  &   660  &   $2.3 \times 10^{-07}$   &  36 \\ 
     200        &  $4.2 \times 10^{-08}$  &   190  &   $9.3 \times 10^{-07}$   &  9\\
     250        &  $3.4 \times 10^{-07}$  &    24  &   $6.7 \times 10^{-06}$   &  2 \\ 
     300        &  $4.4 \times 10^{-06}$  &     2  &   $7.5 \times 10^{-05}$   &  0.2 \\ 
     \hline
    \end{tabular}
    \caption{Isomerisation rate constants}
    \label{tab:rates}
\end{table}

The detection of isotopically labelled compounds can be useful for tracing formation routes; specifically, deuteration has been used in this regard by Bianchi et al. \cite{bianchi} in their study of \ce{CH3C#N} in the SVS13-A Class I hot corino.  Unfortunately, such a tool is unavailable here since the hydrogen transferred has been sourced from the water-ice; however were \ce{CH3C(OD)NH} or \ce{CH3C(OH)ND} to be detected then that might prompt other avenues of investigation.

\begin{figure}
\centering
\includegraphics[width=\textwidth,scale=0.05]{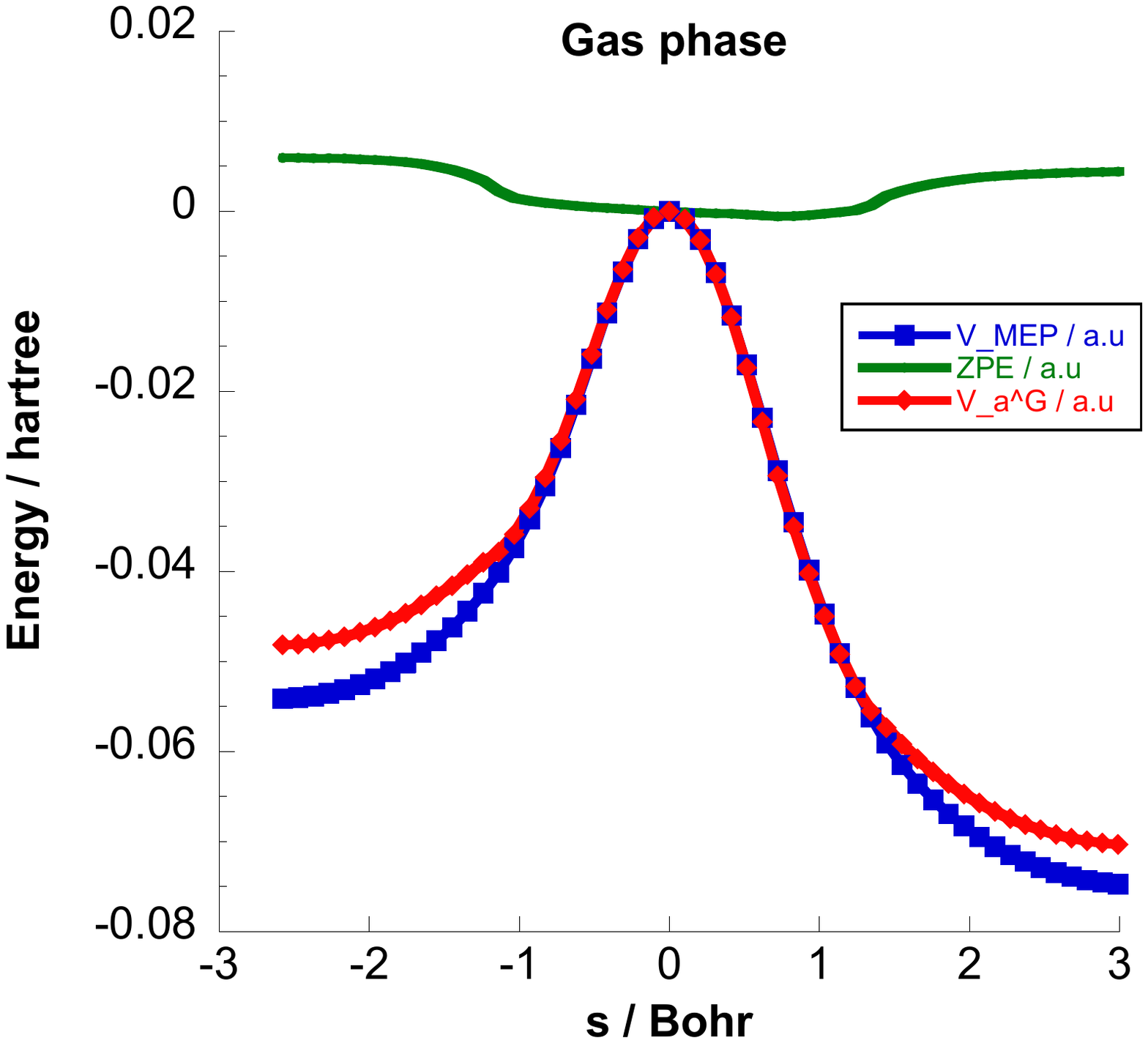}
\caption{Classical potential energies $V_{MEP}$, ground-state vibrational adiabatic potential energy ($V^G_a$), and ZPE as functions of $s$ / Bohr for 1,3[H]-transfer in gas phase}
\label{Figure:7}
\end{figure}

In Fig.~\ref{Figure:7} we plot the potential energy, the zero point energy (ZPE), and vibrationally adiabatic ground state energy $V_a^G$ along the minimum energy path for the gas phase 1,3-H-transfer.

\subsubsection{Intermolecular route}
The same outcome, that is \ce{RC(OH)NH $\to$ RC(O)NH2}, can come about by the attack of hydronium at the N-atom leading similarly to \ce{R-C(OH)=NH2+ + H2O} and attack by a further water molecule at the O-atom leading to deprotonation, resulting in the final products \ce{R-C(O)-NH2 + H3O+}.

\begin{figure} 
\centering
\includegraphics[width=\textwidth,scale=0.05]{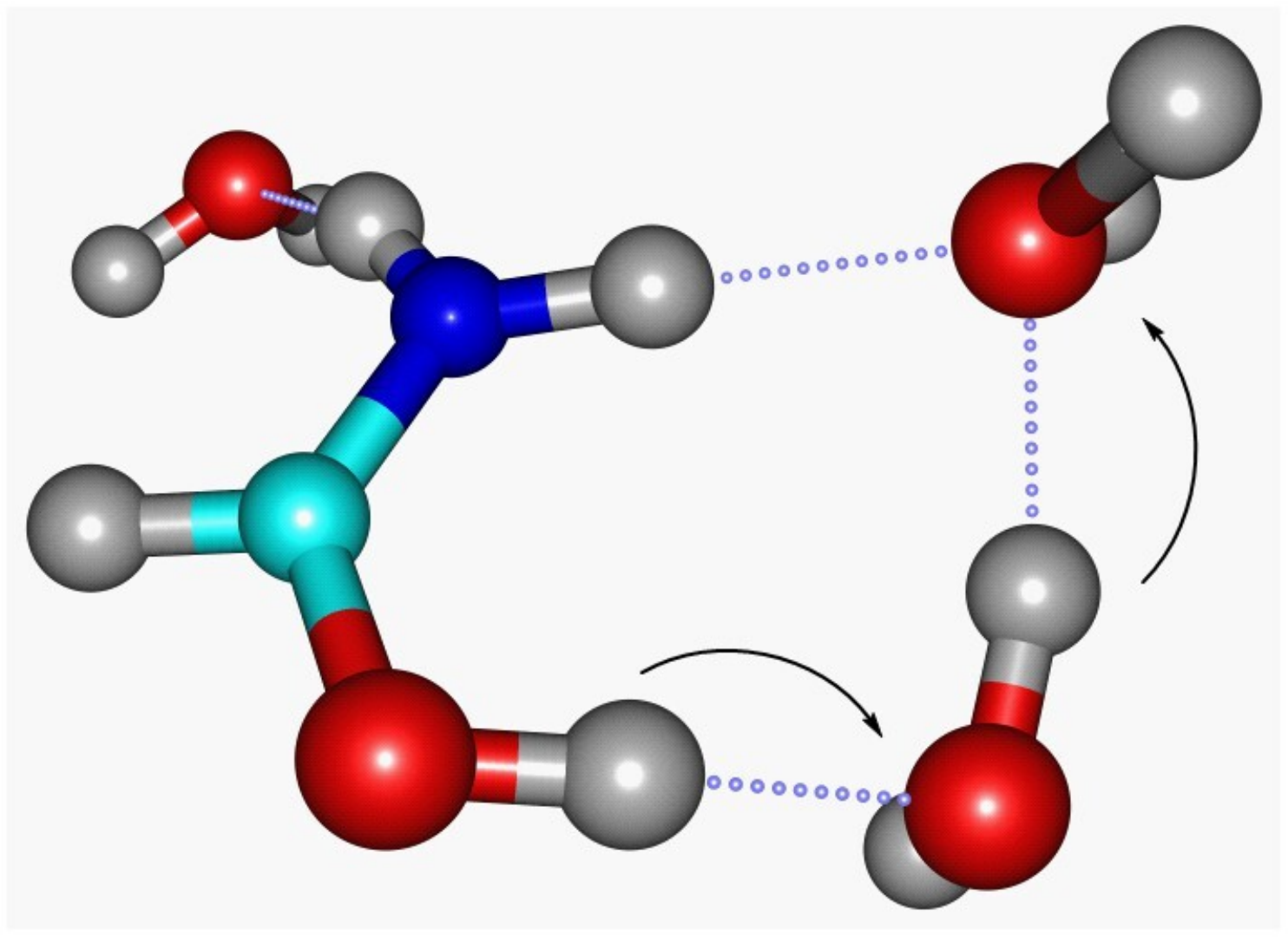}
\caption{Intermolecular \ce{HC(OH)=NH} $\to$ \ce{HC(O)NH2}}
\label{2step}
\end{figure}

So this reaction follows the same course mechanistically as the first; firstly, protonation at the nitrogen atom, which has a tiny barrier of 1.6 kcal/mol (disappears when adding ZPE), is followed by H-abstraction from the OH group and transfer of H to a neighbouring water and transfer of H to a second water, Fig.~\ref{2step}, which just shows the active site.  The reaction can be summarised as:
$$\ce{[HC(OH)NH2+ + 32H2O]} \to \ce{[HC(O)NH2 + 31H2O.H3O+]}$$
and the barrier for this is low at 5.8 kJ mol$^{-1}$. The results are summarised in Table~\ref{table} where $E^{\dagger}$ is the zero-point corrected electronic energy and $\Delta _rH$ is the reaction enthalpy, with both expressed in units of kJ mol$^{-1}$.

\begin{sidewaystable}
\center
\begin{tabular}{lcc}\\ \hline
Reaction                              & $E^{\dagger}$ &  $\Delta _rH$(0 K) \\ \hline
\multicolumn{3}{c}{Initial protonation and 
hydroxy imine formation}\\
\ce{HCN + 32H2O.H3O+ $\to$ HCNH+ + 33H2O $\to$ HC(OH)NH + 31H2O.H3O+}                   &  78.5 & $-82.8$ \\
\ce{CH3CN + 32H2O.H3O+ $\to$
CH3CNH+ + 33H2O $\to$ CH3C(OH)NH + 31H2O.H3O+}                 &  51.9 & $-80.0$ \\
\multicolumn{3}{c}{Intra-molecular amide formation
via 1,3[H]-transfer}\\
\ce{HC(OH)NH (g) $\to$ HC(O)NH2(g)}      & 135.5 & $-57.6$\\
\ce{CH3C(OH)NH (g) $\to$ CH3C(O)NH2 (g)} & 127.6 & $-57.1$\\
\ce{HC(OH)NH + 31H2O.H3O+ $\to$ 
HC(O)NH2 + 31H2O.H3O+}             & 168.8 & $-81.5$\\
\ce{CH3C(OH)NH + 31H2O.H3O+ $\to$ CH3C(O)NH2 + 31H2O.H3O+}         & 141.5 & $-90.1$\\
\multicolumn{3}{c}{Inter-molecular amide formation 
via protonation/deprotonation}\\
\ce{ HC(OH)NH + 31H2O.H3O+$\to$ HC(OH)NH2+ + 32H2O}   & 1.6 &$-75.5$\\
\ce{HC(OH)NH2+ + 32H2O $\to$   HC(O)NH2 + 31H2O.H3O+} & 5.8 &-4.83 \\
\ce{CH3C(OH)NH2+ + 32H2O $\to$ CH3C(O)NH2 + 31H2O.H3O+} &     & \\ \hline
\end{tabular}
\caption{Barrier heights, $E^{\dagger}$, and reaction enthalpies $\Delta _rH$(0~K) / kJ mol$^{-1}$}
\label{table}
\end{sidewaystable}

\section{Discussion}

There are of course other nitriles present in the ISM and one would anticipate that a similar fate would befall them; Manna and Pal \cite{manna} have detected the unfortunately named cyanamide, or amino cyanide \ce{H2NC#N}, in the hot molecular core G10.47+0.03 and outline three possible fates, degradation to \ce{H2N^. + CN^.} by cosmic rays and high-energy photons or by ion--neutral reactions: \ce{H2NC#N + H3+ -> H2NC=NH+ + H2}.  Auto-catalytic addition of water on ice-grains, as per \emph{this work}, would lead to the formation of carbonyl diamide, \ce{OC(NH2)2}, better known as urea, which has been detected previously \cite{belloche}. 

In an extensive computational study by Slate et al. \cite{slate} into urea formation in the ISM they concluded that closed shell reactions had prohibitive barriers but that a route involving charged species was feasible except that their starting point involved iso-cyanic acid \ce{HN=C=O} with protonation at the O-atom followed by addition of ammonia and subsequent deprotonation; a somewhat more complex sequence than that envisaged here but nevertheless comparable.  Previously Brigiano et al. had considered ion--molecule, neutral--neutral and radical reactions leading to the formation of urea but only in the gas-phase \cite{brig}.

\section{Conclusion}\label{sec13}

The auto-catalytic addition of water to \ce{RC#N} triple bonds is shown to be a credible process on water clusters which are impacted by hydrons \ce{H+}.  The high mobility of the hydron through the cluster leads to the initial reactive step, protonation at the N-atom in a facile manner. This effectively transforms a bimolecular reaction into a unimolecular process --- thus removing the `collisional handicap' which hampers all gas-phase reactions in the interstellar medium. 

Subsequent attack by  water at carbon and abstraction of \ce{H+} yields the hydroxy imine \ce{RC(OH)NH} which can then undergo a 1,3[H]-transfer reaction either intra-molecularly or inter-molecularly to the amide \ce{RC(O)NH2}. Quantum-mechanical tunnelling plays a key role in these processes.

The recent review by Woon with its stress on the need to pay more attention to cation--ice reactions is shown to be prescient and his call for experimental confirmation timely. \cite{woon}

%\subsection*{Supplementary information}
%This documents all the salient features of each species mentioned in the principal text.
\subsection*{ORCID}
Bouthe\"ina Kerkeni: \textcolor{blue}{0000-0002-5762-5058}\\
John M. Simmie: \textcolor{blue}{0000-0003-0714-7956}

\subsection*{Acknowledgement}
JMS and BK thank the Irish Centre for High-End Computing, ICHEC, for the provision of computational resources (projects: nuig02, ngche102c, ngche115c)
The assistance by David Ferro-Costas (Univ. Santiago de Compostela), author of Pilgrim, is gratefully acknowledged.


\begin{thebibliography}{99}
\bibitem{kolesnikova} Kolesnikov\'a L, Belloche A, Kouck\'y J, Alonso ER, Garrod RT, Lukov\'a K, Menten KM, M\"uller HSP, Kania P, Urban \u{S}.  
Laboratory rotational spectroscopy of acrylamide and search for acrylamide and propionamide towards Sgr B2(N) with ALMA. A \& A 2021;659:A111
\bibitem{ligterink} Ligterink NFW, Ahmadi A, Luitel B, Coutens A, Calcutt H, Tychoniec L, 
Linnartz H, Jørgensen JK, Garrod RT, Bouwman J. The prebiotic molecular inventory of Serpens SMM1: II. 
The building blocks of peptide chains. ACS Earth Space Chem 2022;6:455-467.
\bibitem{adande} Adande GR, Woolf NJ, Ziurys LM. Observations of Interstellar Formamide: Availability of a Prebiotic Precursor in the Galactic Habitable Zone.
Astrobiol 2013;13:439--453.
\bibitem{mcguire} McGuire BA. 2021 Census of Interstellar, Circumstellar, Extragalactic, Protoplanetary Disk, and Exoplanetary Molecules. Astro J 2022;259:30.
\bibitem{caden} Schuessler C, Remijan A, Xue C, Carder J, Scolati H, McGuire B.
Searching for Propionamide (\ce{C2H5CONH2}) Toward Sagittarius B2 at Centimeter Wavelengths. arXiv:2208.05823

\bibitem{barone} Barone V, Latouche C, Skouteris D, Vazart F, Balucani N, Ceccarelli C, Lefloch B. Gas-phase formation of the prebiotic molecule formamide: insights from new quantum computations. MNRAS 2015;453:L31--L35
\bibitem{skouteris} Skouteris D, Vazart F, Ceccarelli C, Balucani N, Puzzarini C, Barone V. 
New quantum chemical computations of formamide deuteration support gas-phase formation of this prebiotic molecule.
MNRAS 2017;468:L1--L5.
\bibitem{song} Song L, Kästner J. Formation of the prebiotic molecule \ce{NH2CHO} on astronomical amorphous solid water surfaces: accurate tunneling rate calculations. Phys Chem Chem Phys, 2016;18:29278
\bibitem{douglas} Douglas KM, Lucas D, Walsh C, West NA, Blitz MA, Heard DE. The gas-phase reaction of \ce{NH2} with formaldehyde (\\e{CH2O}) is not a source of
formamide (\ce{NH2CHO}) in interstellar environments. arXiv:2208.12658
\bibitem{fedoseev} Fedoseev G, Chang K-J,  van Dishoeck EF, Ioppolo S, Linnartz H. 
Simultaneous hydrogenation and UV-photolysis experiments of NO in CO-rich interstellar ice analogues; linking \ce{HNCO}, \ce{OCN-}, \ce{NH2CHO}, and \ce{NH2OH}. MNRAS 2016;460:4297--4309.
\bibitem{jones} Jones BM, Bennett CJ, Kaiser RI. Mechanistical studies on the production of formamide (\ce{H2NCHO}) within interstellar ice analogs. Astro J 2011;734:78.
\bibitem{thripati}  Thripati S, Ramabhadran RO. Pathways for the Formation of Formamide, a Prebiotic Biomonomer: Metal-Ions in Interstellar Gas-Phase Chemistry. J Phys Chem A 2021;125: 3457--3472
\bibitem{chuang} Chuang KJ, J\"ager C, Krasnokutski SA, Fulvio D, Henning T. Formation of the simplest amide in molecular clouds: formamide (\ce{NH2CHO}) and its derivatives in \ce{H2O}-rich and CO-rich interstellar ice analogs upon VUV irradiation. Astro J 2022; 933:107. arXiv:2206.10470v
\bibitem{codella} Lee C-F, Codella C, Ceccarelli C, L\'opez-Sepulcre A. Stratified Distribution of Organic Molecules at the
Planet-Formation Scale in the HH 212 Disk Atmosphere. arXiv:2208.10693
\bibitem{cordiner} Cordiner MA, Remijan AJ, Boissier J, Milam SN, Mumma MJ, Charnley SB, Paganini L, Villanueva G, Bockel\'ee-Morvan D, Kuan Y-J, et al Mapping the release of volatiles in the inner comae of comets C/2012 F6 (Lemmon) and C/2012 S1 (ISON) using the Atacama large millimeter/submillimeter array. Astrophys J 2014;792:L2

\bibitem{loomis} Loomis RA, Cleeves LI, Oberg KI, Aikawa Y, Bergner J, Furuya K, Guzman VV, Walsh C. The Distribution and Excitation of \ce{CH3CN} in a Solar Nebula Analog. Astro J 2018;859:131.

\bibitem{colzi} Colzi L, Rivilla VM, Beltr\'an MT, Jim\'enez-Serra I, Mininni C, Melosso M, Cesaroni R, Fontani F, Lorenzani A, Sanchez-Monge A, et al. The GUAPOS project ii. A comprehensive study of peptide-like bond molecules. A \& A 2021;653.
\bibitem{kalv} Kalv\={a}ns K, Silsbee K. Icy molecule desorption in interstellar grain collisions. MNRAS 2022;515:785--794.
\bibitem{mini} Minissale M, Aikawa Y, Bergin E, et al. Thermal Desorption of Interstellar Ices: A Review on the Controlling Parameters and Their Implications from Snowlines to Chemical Complexity. CS Earth \& Spac Chem 2022;6:59--630.
\bibitem{darla} Darla N, Sharma D, Sitha S. Formation of Formamide from HCN + \ce{H2O}: A Computational 
Study on the Roles of a Second \ce{H2O} as a Catalyst, as a Spectator, and as a Reactant. J Phys Chem A 2020;124:165--175
\bibitem{rimola} Rimola A, Skouteris D, Balucani N, Ceccarelli C, Enrique-Romero J, Taquet V, Ugliengo P. 
Can Formamide Be Formed on Interstellar Ice? An Atomistic Perspective. ACS Earth Space Chem 2018;2:720--734
\bibitem{bulak} Bulak M, Paardekooper DM, Fedoseev G, Linnartz H. Photolysis of acetonitrile in a water-rich ice as a source of complex organic molecules: 
\ce{CH3CN} and \ce{H2O} : \ce{CH3CN} ices. A \& A 2021;647:A82

\bibitem{simmie} Simmie JM. \ce{C2H5NO} Isomers: From Acetamide to 1,2-Oxazetidine and Beyond. J Phys Chem A 2022;126:924--936.
\bibitem{wootten}  Wootten A, Mangum JG, Turner BE, Bogey M, Boulanger F, Combes F, Encrenaz PJ, Gerin M. Detection of Interstellar \ce{H3O+}: A Confirming Line. Astrophys J 1991;380:L79.
\bibitem{tak} Tak FD, Aalto S, Meijerink R. Detection of extragalactic \ce{H3O+}. A \& A 2007;477:5--8.
\bibitem{moon}  Moon ES, Kang H, Oba Y, Watanabe N, Kouchi A. Direct Evidence for Ammonium Ion Formation in Ice Through
Ultraviolet-Induced Acid Base Reaction of \ce{NH3} with \ce{H3O+}. Astrophys J 2010;713:906.



\bibitem{lee} Lee DH, Kang H. Proton Transport and Related Chemical Processes of Ice. J Chem Phys B 2021;125:8270--8281
\bibitem{woon}	Woon DE. Quantum chemical cluster studies of cation-ice reactions for astrochemical applications: Seeking experimental confirmation. Accts Chem Res 2021;54:490-497.
\bibitem{martinez} Martinez R, Agnihotri AN, Boduch P, Domaracka A, Fulvio D, Palumbo ME, Rothard H, Strazzulla G. Production of Hydronium Ion (\ce{H3O+}) and Protonated Water Clusters (\ce{H2O_n.H+}) after Energetic Bombardment  of Water Ice in Astrophysical Environments. J Phys Chem A 2019; 123: 8001--8008.
\bibitem{ziurys} Ziurys LM, Turner BE. Detection of interstellar vibrationally excited HCN. Ap J 1986;302:L31
\bibitem{schilke} Schilke P, Walmsley CM, Henkel C, Millar TJ. Protonated HCN in molecular clouds. A \& A 1991;247:487.
\bibitem{quenard} Qu\'enard D, Vastel C, Ceccarelli C, Hily-Blant P, Lefloch B, Bachiller R. Detection of the \ce{HC3NH+} and \ce{HCNH+} ions in the L1544 pre-stellar core. MNRAS 2017;470:3194
\bibitem{titan} Cravens TE, Robertson IP, Waite JH, Yelle RV, Kasprzak WT, Keller CN,
Ledvina SA, Nieman HB, Luhmann JG, McNutt RL, Ip W-H, de la Haye
V, Mueller-Wodarg I, Wahlund J-E, Anicich VG, Vuitton V.
Composition of Titan’s ionosphere. Geophys Res Lett 2006;33:L07105
\bibitem{fontani} Fontani F, Colzi L, Redaelli E, Sipil\"a O, Caselli P. First survey of \ce{HCNH+} in high-mass star-forming cloud cores. A \& A 201;651:A94.
\bibitem{mari} Marimuthu AN, Huis in’t Veld F, Thorwirth S, Redlich B, Br\"unken S.
Infrared predissociation spectroscopy of protonated methyl cyanide, \ce{CH3CNH+}. J Mol Spectros 2021;379:
\bibitem{hops} Lee DH, Choi CH, Choi TH, Sung BJ, Kang H. Asymmetric transport mechanisms of hydronium and hydroxide ions in amorphous solid water: Hydroxide goes Brownian while hydronium hops. J Phys Chem Letts 2014;5:2568-2572.
\bibitem{krasno}	Krasnokutski SA, Chuang KJ, Jager C, Ueberschaar N, Henning T. A pathway to peptides in space through the condensation of atomic carbon. Nature Astronomy 2022;6:381--386.
\bibitem{canepa} Canepa C. A Model Study on the Dynamics of the Amino Acid Content in Micrometeoroids during Atmospheric Entry.
Chemistry 2020;2:918--936.


\bibitem{gauss} Gaussian 16, Revision C.01, Frisch, M. J.; Trucks, G. W.; Schlegel, H. B.; Scuseria, G. E.; Robb, M. A.; Cheeseman, J. R.; Scalmani, G.; Barone, V.; Petersson, G. A.; Nakatsuji, H.; Li, X.; Caricato, M.; Marenich, A. V.; Bloino, J.; Janesko, B. G.; Gomperts, R.; Mennucci, B.; Hratchian, H. P.; Ortiz, J. V.; Izmaylov, A. F.; Sonnenberg, J. L.; Williams-Young, D.; Ding, F.; Lipparini, F.; Egidi, F.; Goings, J.; Peng, B.; Petrone, A.; Henderson, T.; Ranasinghe, D.; Zakrzewski, V. G.; Gao, J.; Rega, N.; Zheng, G.; Liang, W.; Hada, M.; Ehara, M.; Toyota, K.; Fukuda, R.; Hasegawa, J.; Ishida, M.; Nakajima, T.; Honda, Y.; Kitao, O.; Nakai, H.; Vreven, T.; Throssell, K.; Montgomery, J. A., Jr.; Peralta, J. E.; Ogliaro, F.; Bearpark, M. J.; Heyd, J. J.; Brothers, E. N.; Kudin, K. N.; Staroverov, V. N.; Keith, T. A.; Kobayashi, R.; Normand, J.; Raghavachari, K.; Rendell, A. P.; Burant, J. C.; Iyengar, S. S.; Tomasi, J.; Cossi, M.; Millam, J. M.; Klene, M.; Adamo, C.; Cammi, R.; Ochterski, J. W.; Martin, R. L.; Morokuma, K.; Farkas, O.; Foresman, J. B.; Fox, D. J. Gaussian, Inc., Wallingford CT, 2016.
\bibitem{wB97} Chai J-D, Head-Gordon M. Long-range corrected hybrid density functionals with damped atom–atom dispersion corrections. PCCP 2008;10:6615--6620.

\bibitem{Hratchian2004} Hratchian HP, Schlegel HB. Accurate reaction paths using a Hessian
based predictor–corrector integrator.  J Chem Phys 2004;120:9918.
\bibitem{Hratchian2005} Hratchian HP, Schlegel HB. Using Hessian Updating To Increase the Efficiency of a Hessian Based Predictor-Corrector Reaction Path Following Method. JCTC 2005;1:61--69.
\bibitem{page} Page M, McIver Jr JW. On evaluating the reaction path Hamiltonian. J Chem Phys 1988;88:922--935.

\bibitem{sct} Liu Y-P, Lynch GC, Truong TN, Lu D-H, Truhlar DG. Molecular Modelling of the Kinetic Isotoe Effect for the [1,5]-Sigmatropic Rearrangement of cis-1,3-Pentadiene. JACS 1993;115:2408.

\bibitem{qrc} Wonchoba SE, Hu W-P, Truhlar DG. Theoretical and Computational Approaches to Interface Phenomena, eds. Sellers HL, Golab JT. Plenum, New York, 1994. Pp. 1 Reaction Path Approach to Dynamics at a Gas-Solid Interface: Quantum Tunnelling Effects for an Adatom on a Non-Rigid Metallic Surface.
\bibitem{pil} PILGRIM v2021.5 \verb|https://github.com/cathedralpkg/pilgrim|
\bibitem{Rim} Enrique-Romero J, Rimola A, Ceccarelli C, Ugliengo P, Balucani N, Skouteris D. 
Reactivity of HCO with \ce{CH3} and \ce{NH2} on Water Ice Surfaces. A Comprehensive Accurate Quantum Chemistry Study.
ACS Earth Space Chem 2019;3:2158--2170
\bibitem{hunter} Hunter EP, Lias SG. Evaluated Gas Phase Basicities and Proton Affinities of Molecules: An Update, J Phys Chem Ref Data, 1998; 27: 413--656.
\bibitem{pilgrim} Ferro-Costas D, Truhlar DG, Fern\'andez-Ramos A. Pilgrim: A thermal rate constant calculator and a chemical kinetics simulator. Comp Phys Comms 2020;256:107457

\bibitem{con} Concepci\'on JG, Jim\'enez-Serra I, Corchado JC, Rivilla VM, Mart\'{\i}n-Pintado J. The Origin of the E/Z Isomer Ratio of Imines in the Interstellar Medium.
Ap J L 2021;912:L6
\bibitem{bianchi} Bianchi E, Ceccarelli C, Codella C, SOLIS XV. \ce{CH3CN} deuteration in the Class I hot corino. A \& A 2022; 662:A103.
\bibitem{manna} Manna A, Pal S.  Detection of interstellar cyanamide (\ce{NH2CN}) towards the hot molecular core G10.47+0.03. arXiv:2207.04877

%\bibitem{bulak} Bulak M, Paardekooper DM, Fedoseev G, Linnartz H. Astro \& Astrophys 2021;647:A82
\bibitem{belloche} Belloche A, Garrod RT, Müller HSP, Menten KM, Medvedev I, Thomas J, Kisiel Z, Re-exploring Molecular Complexity with ALMA (ReMoCA): interstellar detection of urea. A \& A;2019:628,A10. \emph{See also}  A \& A 2020;637:C4.
\bibitem{slate} Slate ECS, Barker R, Euesden RT, Revels MR, Meijer AJHM. Computational studies into urea formation in the interstellar medium. MNRAS 2020;497:5413--5420
\bibitem{brig} Brigiano FS, Jeanvoine Y, Largo A, Spezia R.The formation of urea in space I. Ion-molecule, neutral-neutral, and radical gas-phase reactions. A \& A 2018;610:A26.
\end{thebibliography}
\end{document}